\documentclass[11pt, a4paper]{article}

\usepackage[a4paper, margin=2.5cm]{geometry}
\usepackage[utf8]{inputenc}
\usepackage[T1]{fontenc}
\usepackage{microtype}

\usepackage{amsmath,amssymb}
\usepackage{booktabs}
\usepackage{tabularx}
\usepackage{array}
\usepackage{cite}
\usepackage{hyperref}
\usepackage{graphicx}

\hypersetup{
    colorlinks=true,
    linkcolor=blue,
    citecolor=blue,
    urlcolor=blue
}

\title{\textbf{Recurrent Bipartite Graph Neural Networks for QC-MDPC Error Decoding in McEliece Cryptosystems}}

\author{
    Parth Rajeshkumar Tandel$^{1,*}$ \and Elisabeth Fink$^{1,\dagger}$ \\[1em]
    \small $^{1}$Middlesex University, Department of Computer Science, Hendon, London NW4 4BT, UK \\
    \small Corresponding authors: $^{*}$\texttt{ptandel.de@gmail.com}, $^{\dagger}$\texttt{e.fink@mdx.ac.uk}
}

\date{}

\begin{document}

\maketitle

\begin{abstract}
As the advent of large-scale quantum computing threatens classical encryption, Quasi-Cyclic Moderate Density Parity-Check (QC-MDPC) codes have emerged as a foundational element of Post-Quantum Cryptography (PQC). However, decoding these relatively dense codes using classical Belief Propagation algorithms often leads to failures caused by short graph cycles that stall the decoder. In cryptographic applications, these failures are severe security vulnerabilities that leak secret key information. To address this, this paper proposes a Recurrent Bipartite Graph Neural Network (R-BGNN) as proof of concept. By integrating Gated Recurrent Units (GRUs) into the message-passing framework and employing a multi-topology curriculum training strategy, the R-BGNN actively learns to break limit-cycle oscillations. Evaluated across 20 unseen matrix topologies, the proposed architecture significantly outperforms classical Bit-Flipping and Min-Sum baselines, reducing the Decoding Failure Rate (DFR) by 63.7\% at moderate noise levels ($t=10$) while converging in fewer iterations.
\end{abstract}


\section{Introduction}
Public key encryption (PKE) was a breakthrough in the field of modern cryptography, majorly relying on hard mathematical problems such as prime factorization and discrete logarithms. However, algorithms designed for large-scale quantum computers possess the potential to solve these mathematical defenses in polynomial time, rendering current encryption protocols at threat \cite{ref1}. This anticipated future vulnerability, termed `Quantum D-Day', will render modern public-key standards such as RSA and Elliptic Curve Cryptography (ECC) completely insecure, compelling an urgent global transition toward Post-Quantum Cryptography (PQC).

In response, lattice-based, hash-based and code-based cryptography have emerged as highly promising fields within PQC \cite{ref2}. Among these, the McEliece public-key cryptosystem stands out as one of the most thoroughly cryptanalyzed candidates \cite{ref3}. However, traditional McEliece variants utilizing Goppa codes faced significant deployment barriers due to their requirement for very large public keys. To overcome this limitation, \cite{ref4} introduced the Quasi-Cyclic Moderate Density Parity-Check (QC-MDPC) McEliece variant, which relies on MDPC codes to provide highly compact keys. By adopting a quasi-cyclic (QC) structure, each circulant block of the parity-check matrix can be compactly represented by its first row, significantly reducing memory overhead and positioning QC-MDPC codes as a primary foundation for National Institute of Standards and Technology (NIST) PQC finalists like the Bit-Flipping Key Encapsulation (BIKE) cryptosystem \cite{ref5}.

\subsection{PQC \& QC-MDPC Codes}
The structural paradigm of QC-MDPC codes diverges entirely from that of standard telecommunication Low-Density Parity-Check (LDPC) codes \cite{ref6}. LDPC codes rely on extreme sparsity to facilitate optimal iterative decoding; however, utilizing highly sparse matrices in cryptographic contexts leaves the system vulnerable to attacks \cite{ref4}. To maintain cryptographic uniformity, QC-MDPC parity-check matrices are intentionally built from random, large circulant blocks with a much higher density $W = \mathcal{O}(\sqrt{N})$ than standard communication codes \cite{ref7}. While this increased density ensures resistance against structural attacks, it introduces severe topological complications. Most notably, the increased row weights cause an unavoidable abundance of 4-cycles to grow throughout the Tanner graph representations which are bipartite graphs connecting variable bit nodes to parity-check nodes of these matrices \cite{ref5, ref8}. As literature demonstrates, these structural anomalies shift the primary challenge of QC-MDPC cryptography from code generation to reliable message decoding.

\subsection{Classical BP Decoding \& Its Failure Points}
To decipher encrypted messages, QC-MDPC systems utilize iterative decoding algorithms such as Belief Propagation (BP) or Bit-Flipping \cite{ref9, ref10}. In standard implementations, exact BP requires heavy computational complexity due to the use of hyperbolic tangent functions \cite{ref10}. To reduce this, the Min-Sum algorithm simplifies node updates into basic sign-multiplication and minimum-value comparisons \cite{ref11}. However, this simplification comes at a severe cost: classical Min-Sum relies on over-approximations that drastically overestimate Log-Likelihood Ratio (LLR) magnitudes, degrading the error-correction capability of the decoder \cite{ref10, ref11}. Because short, dense codes are plagued by harmful short cycles, the fundamental assumption of BP, that incoming messages are statistically independent, is violated \cite{ref12}. These short cycles give rise to dense ``trapping sets'' that artificially correlate messages and degrade LLR updates across iterative decoding rounds \cite{ref12}.

Classical attempts to resolve this have proven computationally burdensome. For instance, \cite{ref13} proposed Multiple-Bases Belief-Propagation (MBBP), which attempts to decode high-rate algebraic block codes which are error-correcting codes that divide data into fixed-length blocks of bits and add redundant parity bits to detect and correct errors, by running multiple BP decoders in parallel over structurally diverse parity-check matrix permutations; however, this approach linearly multiplies hardware complexity. More recently, \cite{ref14} proposed modifying the Tanner graph itself by constructing Low-Weight Parity-Check (LWPC) matrices and inserting auxiliary variable nodes (AVNs) alongside redundant check nodes to physically break 4-cycles. While effective at cycle elimination, introducing AVNs, which lack direct soft channel observations, can weaken cycle connectivity and unacceptably expand graph dimensions, thereby throttling expected performance gains \cite{ref12}.

Consequently, iterative decoding on QC-MDPC graphs exhibits an alarming ``error floor'' behavior where Decoding Failure Rates (DFR) stall at unacceptable levels as block length increases \cite{ref15}. In cryptographic applications, this decoder failure is not merely a reliability issue, but a severe security flaw. \cite{ref16} introduced the reaction attack mechanics, which was later designated as the ``GJS Attack'' in subsequent literature \cite{ref5, ref15}, wherein an adversary systematically injects specific error patterns and observes whether the target's decoder fails or succeeds. Because decoding failures correlate strongly with the distance between non-zero entries in the secret parity-check matrix, an attacker can use these failure reactions to construct a ``distance spectrum'' and entirely reconstruct the private key \cite{ref16}. Therefore, lowering the DFR is paramount to defending against reaction-based key-recovery attacks and ensuring Chosen-Ciphertext Attack (IND-CCA2) security.

\subsection{Deep Unrolled Neural Decoders}
To overcome the catastrophic error floors of classical iterative decoding without manually altering the graph topology, researchers have increasingly turned to deep learning frameworks. The pioneering approach in this domain is ``deep unrolling,'' which maps the discrete, iterative steps of the BP algorithm directly into the hidden layers of a feed-forward neural network \cite{ref17}. By assigning trainable weights to the edges of the Tanner graph, this architecture creates a ``soft'' Tanner graph capable of actively scaling and damping messages to compensate for the detrimental effects of short cycles \cite{ref17}.

A critical structural requirement of these unrolled networks is the preservation of message-passing symmetry \cite{ref17}. Mathematically, this prevents the neural decoder from developing a bias toward specific bit patterns and drastically simplifies network training by allowing the model to be trained entirely on the all-zero codeword rather than an exponentially large dataset of $2^k$ messages \cite{ref17}.

Despite their success in improving error-correction thresholds, early unrolled neural networks introduced significant engineering bottlenecks. Assigning individual weights to every edge of a dense MDPC graph leads to massive parameter explosion, making the models exceptionally difficult to train. Recent literature has addressed this by introducing shared-weight architectures, such as the Shared Neural Normalized Min-Sum (SNNMS) decoder, which enforces the sharing of correction factors across all edges within the same iteration layer, where a single decoding iteration represents one complete round of bipartite message passing comprising check-to-variable (C2V) and variable-to-check (V2C) message updates across the Tanner graph, reducing parameter complexity for a Tanner graph of $E$ edges and $L_{\max}$ number of maximum iterations, from $\mathcal{O}(E \times L_{\max})$ to $\mathcal{O}(L_{\max})$ and drastically improving training stability \cite{ref10}. Nevertheless, a major limitation persists within the deep unrolling paradigm: standard unrolled neural networks lack topological scalability. Because the neural architecture is hard-coded to a specific Tanner graph, the model cannot seamlessly scale to varying code lengths or rates without undergoing computationally expensive retraining from scratch \cite{ref18}.

\subsection{Graph Neural Networks (GNNs) for Decoding}
To resolve the topological rigidity and scalability limitations of deep unrolled networks, contemporary research has pivoted towards Graph Neural Networks (GNNs). They abstract the decoding process into generalized graph operations that naturally achieve ``algorithmic alignment'' with the message-passing structure of BP on bipartite Tanner graphs \cite{ref18}. Rather than relying on rigid, handcrafted update formulas, GNN decoders treat channel error correction as a node-classification task on the graph, dynamically learning adaptive message-passing rules that handle loopy topologies and short-cycle trapping structures far more effectively than standard BP \cite{ref19, ref20}.

Edge-Weighted GNNs (EW-GNN) have proven highly effective for short block codes. By employing a feed-forward neural network to compute multiplicative edge weights dynamically based on local neighborhood message reliabilities, EW-GNNs suppress unreliable messages generated by short cycles \cite{ref18}. Crucially, because the neural weight generator works on local node features rather than fixed Tanner graph positions, the parameter count remains completely independent of codeword length ($\mathcal{O}(1)$ constant parameter footprint). This allows the model trained on a smaller code topology (e.g., $n=32$) to be applied directly to a larger, more complex code (e.g., $n=128, 256$) without any parameter adjustment \cite{ref18}.

Further bridging the gap between classical theory and deep learning, hybrid models such as Neural Enhanced Belief Propagation (NEBP) combine mathematically rigorous BP updates with GNN-generated message refinements at each iteration \cite{ref21}. By operating conjointly with classical algorithms, NEBP leverages the adaptive learning capability of GNNs to handle complex, non-Gaussian, or bursty noise profiles where standard BP fails to converge \cite{ref21}. Recent studies by \cite{ref20} demonstrate that integrating attention-weighted message aggregation and skip connections into graph decoders accelerates convergence, requiring significantly fewer iterations than classical BP while maintaining superior bit error rates across dynamic channel conditions. Complementing these decoder-focused developments, differentiable tensor reformulations of BP now allow data-driven optimization of the factor graph itself, enabling researchers to gradient-optimize the binary parity-check matrix $H$ directly under channel noise simulations \cite{ref22}. Table~\ref{tab:summary_architectures} summarizes the complexity, scalability and applications of some of the architectures cited in this paper.
% Add these to your document preamble if not already present:
% \usepackage{tabularx}
% \usepackage{booktabs}
% \usepackage{array}

\begin{table*}[t]
\centering
\small
\caption{Summary of various architectures}
\label{tab:summary_architectures}
\renewcommand{\arraystretch}{1.4} % Replicates the open vertical spacing from the Word table

\begin{tabularx}{\textwidth}{
    >{\hsize=0.85\hsize\raggedright\arraybackslash}X   % Architecture (Bold, left-aligned)
    >{\centering\arraybackslash}p{1.6cm}               % Literature (Fixed width, centered to match brackets)
    >{\hsize=1.25\hsize\raggedright\arraybackslash}X   % Weight Granularity (Wide, left-aligned)
    >{\hsize=1.00\hsize\raggedright\arraybackslash}X   % Scalability (Medium, left-aligned)
    >{\hsize=0.75\hsize\raggedright\arraybackslash}X   % Secret Key (Narrow, left-aligned)
    >{\hsize=1.15\hsize\raggedright\arraybackslash}X   % Primary Application (Wide, left-aligned)
}
\toprule
\textbf{Architecture} & \textbf{Literature} & \textbf{Weight Granularity \& Parameter Complexity} & \textbf{Scalability across Lengths $n$} & \textbf{Secret Key Graph Preserved?} & \textbf{Primary Application Domain} \\
\midrule
\textbf{Deep Unrolled BP} & \cite{ref17} & Per-edge per-layer: $\mathcal{O}(E \times L_{\max})$ & No (Hard-coded to fixed graph) & Yes & High-Density Algebraic Block Codes \\

\textbf{Attentive SNNMS (A-SNNMS)} & \cite{ref10} & Per-layer shared: $\mathcal{O}(L_{\max})$ & No (Requires retraining per code size) & Yes & Telecommunication LDPC (5G / Wi-Fi 6) \\

\textbf{Edge-Weighted GNN (EW-GNN)} & \cite{ref18} & MLP edge weight generator: $\mathcal{O}(1)$ constant & Yes (Transferable across $n$) & Yes & Short Block Communication Codes \\

\textbf{Neural Enhanced BP (NEBP)} & \cite{ref21} & Factor Graph GNN refinement: $\mathcal{O}(GNN)$ & Partial (Graph-dependent) & Yes & Non-Gaussian / Bursty Noise Channels \\

\textbf{Differentiable Factor Graph} & \cite{ref22} & Gradient optimization of binary matrix $H$ & N/A (Code design tool) & Alters matrix $H$ & Differentiable Short Code Design \\
\bottomrule
\end{tabularx}
\end{table*}

\subsection{Factor Graph Optimization \& The Research Gap}
While the integration of GNNs into channel decoding represents a significant leap forward, critical deficiencies remain when applying these models to the specific constraints of PQC. Literature reveals that static message passing in both classical and early neural decoders ignores real-time reliability differences across iterations \cite{ref10}. Consequently, standard iterative decoders structurally fail when encountering near-codeword absorbing states, where the syndrome resulting from the decoding is of very low Hamming weight \cite{ref15} and dense 4-cycle graph topologies \cite{ref5} inherently present in QC-MDPC matrices.

As established, classical workarounds like MBBP parallel decoders \cite{ref13} and AVN graph expansions \cite{ref14} introduce high hardware overhead or alter key dimensions. Furthermore, while gradient-based optimization over factor graphs has proven capable of discovering superior binary parity-check matrix structures \cite{ref22}, standard unrolled neural decoders operate purely as feed-forward architectures, lacking the recurrent memory states necessary to track temporal message dynamics or detect when decoding has stalled in an absorbing state.

Therefore, a distinct research gap emerges at the intersection of deep learning and PQC. There is an urgent need to combine unrolled graphical message passing with recurrent memory mechanisms such as Gated Recurrent Units (GRUs) \cite{ref23}, to actively retain prior decoding states, dynamically break limit cycles in PQC matrices, and suppress the DFR required to prevent reaction-based key-recovery attacks, all without artificially altering the underlying Tanner graph.

To address this gap, this paper introduces a Recurrent Bipartite Graph Neural Network (R-BGNN) for QC-MDPC decoding in McEliece-style cryptosystems by embedding GRUs directly into the check and variable nodes of the Tanner graph to provide temporal memory across unrolled iterations. To ensure the model generalizes well to new keys, we use topological degree normalization and train the network on dynamically sampled matrix topologies using a curriculum learning approach. Finally, we evaluate the model's accuracy and convergence speed across unseen master key topologies under exact cryptographic error weights, comparing its performance against classical Bit-Flipping and Min-Sum BP decoders.

\section{Methods}
The methodology encompasses five primary phases: dataset generation, formulation of classical mathematical baselines, design of the R-BGNN, implementation of the training loop, and execution of an evaluation protocol.

\subsection{The QC-MDPC Decoding Process}
To establish a clear foundational framework for evaluating neural and classical decoders, it is necessary to formalize the mechanics of syndrome decoding on a bipartite graph. In McEliece-style code-based cryptosystems, an encrypted ciphertext $x \in \mathbb{F}_2^n$ is generated by encoding a plaintext message $m \in \mathbb{F}_2^{n-r}$ using a public systematic generator matrix $G \in \mathbb{F}_2^{(n-r) \times n}$ and adding a random binary noise vector $e \in \mathbb{F}_2^n$ of Hamming weight $t$, such that $x = m G \oplus e$ \cite{ref4, ref9}.

At the receiver, the secret parity-check matrix $H \in \mathbb{F}_2^{r \times n}$ is used to isolate the noise from the message by computing the syndrome vector $s \in \mathbb{F}_2^r$:
\begin{equation}
s = H \cdot x^T = H \cdot (m G \oplus e)^T = H \cdot e^T
\end{equation}
Because $m G H^T = 0$, the syndrome vector $s$ depends strictly on the error vector $e$ and the secret matrix $H$ \cite{ref9}. Each entry $s_j \in \{0,1\}$ in the syndrome vector indicates whether the $j$-th parity-check constraint is satisfied ($s_j = 0$) or unsatisfied ($s_j = 1$). Decoding is complete when the estimated error vector $\hat{e}$ satisfies $H \hat{e}^T = s$, yielding a zero residual syndrome $s = \mathbf{0}_r$.

This decoding process is mapped onto a bipartite Tanner graph $\mathcal{G} = (\mathcal{V}, \mathcal{C}, \mathcal{E})$, where $n$ variable nodes $v_i \in \mathcal{V}$ correspond to codeword bit positions, $r$ check nodes $c_j \in \mathcal{C}$ correspond to parity-check equations, and an edge $e_{j,i} \in \mathcal{E}$ connects check node $c_j$ to variable node $v_i$ if and only if $H_{j,i} = 1$ \cite{ref8, ref18}.

Within this graphical framework, a single decoding iteration represents one complete two-phase round of message passing across all edges in the Tanner graph \cite{ref8, ref24}. Let $\mathcal{M}_i$ denote the set of check nodes connected to variable node $v_i$, and let $\mathcal{N}_j$ denote the set of variable nodes connected to check node $c_j$. At each iteration $t$, extrinsic reliability messages in the form of LLRs are exchanged across connected edges via two distinct phases:

\paragraph{Check-to-Variable (C2V) Update:} Each check node $c_j$ computes the extrinsic check-to-variable message $L_{c_j \to v_i}^{(t)}$ transmitted back to variable node $v_i$ by evaluating parity constraints across all connected variable nodes except $v_i$:
\begin{equation}
L_{c_j \to v_i}^{(t)} = 2 \cdot \tanh^{-1} \left( \prod_{v_k \in \mathcal{N}_j \setminus \{v_i\}} \tanh \left( \frac{L_{v_k \to c_j}^{(t-1)}}{2} \right) \right)
\end{equation}
Here, $\mathcal{N}_j \setminus \{v_i\}$ denotes the set of all variable nodes connected to check node $c_j$, excluding the recipient node $v_i$.

\paragraph{Variable-to-Check (V2C) Update:} Each variable node $v_i$ aggregates its intrinsic initial channel observation $L_i^{(0)}$ with incoming extrinsic messages from all connected check nodes except $c_j$ to produce the outgoing variable-to-check message $L_{v_i \to c_j}^{(t)}$:
\begin{equation}
L_{v_i \to c_j}^{(t)} = L_i^{(0)} + \sum_{c_m \in \mathcal{M}_i \setminus \{c_j\}} L_{c_m \to v_i}^{(t)}
\end{equation}
Here, $\mathcal{M}_i \setminus \{c_j\}$ represents the set of all check nodes connected to variable node $v_i$, excluding the target check node $c_j$.

At the conclusion of iteration $t$, the total posterior LLR for each bit is computed by summing the intrinsic observation with all incoming check messages:
\begin{equation}
L_i^{(t)} = L_i^{(0)} + \sum_{c_j \in \mathcal{M}_i} L_{c_j \to v_i}^{(t)}
\end{equation}
A hard-decision bit estimate $\hat{e}_i^{(t)}$ is produced via:
\begin{equation}
\hat{e}_i^{(t)} = \begin{cases}
1, & \text{if } L_i^{(t)} < 0 \\
0, & \text{if } L_i^{(t)} \ge 0
\end{cases}
\end{equation}
If the resulting error vector satisfies $H \cdot (\hat{e}^{(t)})^T = s$, decoding terminates; otherwise, message passing advances to iteration $t+1$ up to $T_{\max}$ iterations.

Defining an iteration as this complete C2V and V2C exchange provides a standardized baseline for comparing the convergence speed and parameter efficiency of feed-forward, unrolled, and recurrent decoders.

\subsection{Dataset Generation}
The foundational step of the research framework involves generating QC-MDPC matrices used by McEliece-style PQC candidates \cite{ref4}. A QC-MDPC code is defined by its block index $n_0$, circulant block size $r$, and uniform row weight $W$. For this paper, the framework generates codes with parameters $n_0 = 2$, block size $r = 127$, and total block length $N = n_0 \times r = 254$.

\begin{figure}[htbp]
\centering
\includegraphics[width=0.85\linewidth]{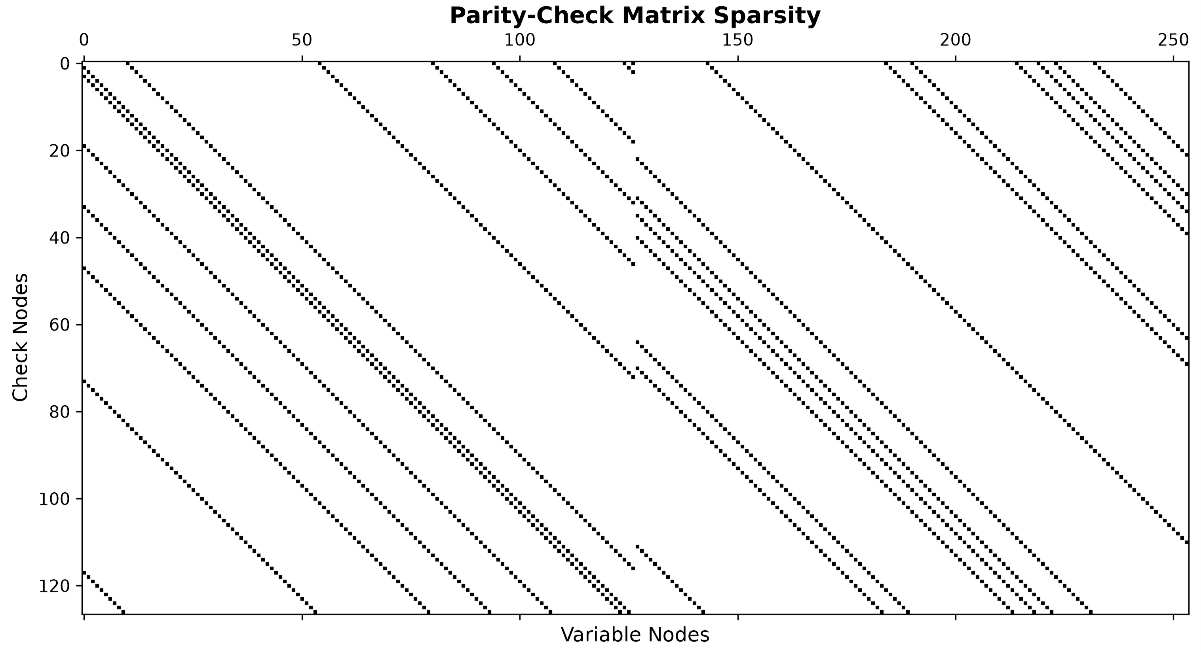} % Replace 'your_filename.png' with your actual file name
\caption{A random sample binary Parity-Check Matrix, $N=254, r=127, W=14, d=7$ (black dots represent set bits)}
\label{fig:matrix_sparsity}
\end{figure}

The parity-check matrix $H = [H_0 \mid H_1] \in \mathbb{F}_2^{r \times N}$ is constructed by generating two random, sparse circulant blocks $H_0 \in \mathbb{F}_2^{r \times r}$ and $H_1 \in \mathbb{F}_2^{r \times r}$ \cite{ref4}. The first row of each block is constructed by randomly allocating non-zero entries such that the row weight of each block equals $W/2 = 7$, yielding a total row weight of $W = 14$. Subsequent rows are formed via cyclic right-shifts of the preceding row. This structure ensures a uniform row degree of $W = 14$ for all check nodes and a uniform column degree of $d = 7$ for all variable nodes \cite{ref7}. Figure~\ref{fig:matrix_sparsity} shows the structure of one such matrix generated using these specifications.

% \begin{figure}[htbp]
% \centering
% \framebox{\parbox{0.85\linewidth}{\centering\vspace{1.5cm}\textbf{Figure 1: Parity-Check Matrix Sparsity}\\\small Binary Parity-Check Matrix: $N=254, r=127, W=14, d=7$ (black dots represent set bits)\vspace{1.5cm}}}
% \caption{A random sample binary Parity-Check Matrix, $N=254, r=127, W=14, d=7$ (black dots represent set bits)}
% \label{fig:matrix_sparsity}
% \end{figure}

Because cryptographic security dictates that these matrices remain relatively dense ($W = \mathcal{O}(\sqrt{N})$) compared to telecommunication LDPC codes ($W = \mathcal{O}(1)$), they inherently generate loopy graph structures \cite{ref4, ref7}. To quantify this, the framework executes a topological profiling phase upon matrix generation. The profiling engine calculates the check-node intersection spectrum by analyzing the inner product matrix $A = H H^T \in \mathbb{Z}^{r \times r}$. The off-diagonal entries $A_{i,j}$ (for $i \neq j$) denote the number of shared variable nodes between check nodes $c_i$ and $c_j$; an overlap of $A_{i,j} \ge 2$ mathematically guarantees the existence of a length-4 cycle (4-cycle) in the bipartite Tanner graph \cite{ref5}.

\begin{figure}[htbp]
\centering
\includegraphics[width=0.85\linewidth]{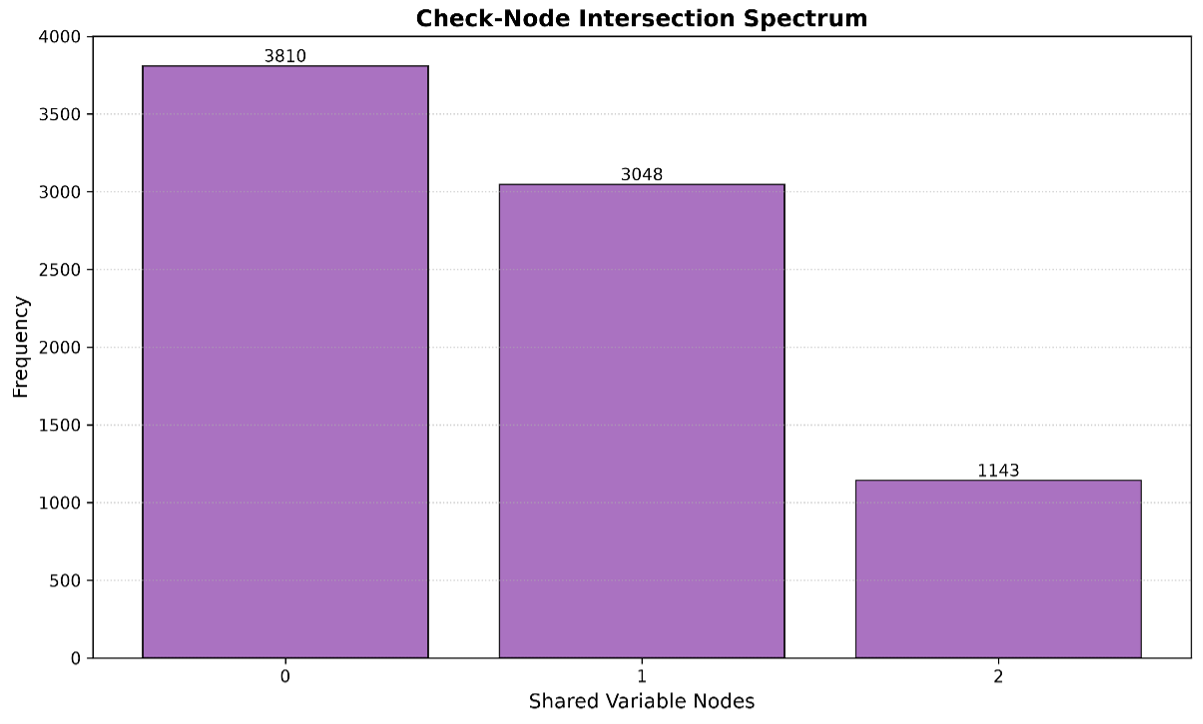} 
\caption{Check-Node Intersection Frequency plot showing 1000+ intersections of more than two shared Variable Nodes}
\label{fig:intersection_spectrum}
\end{figure}

\begin{figure}[htbp]
\centering
\includegraphics[width=0.85\linewidth]{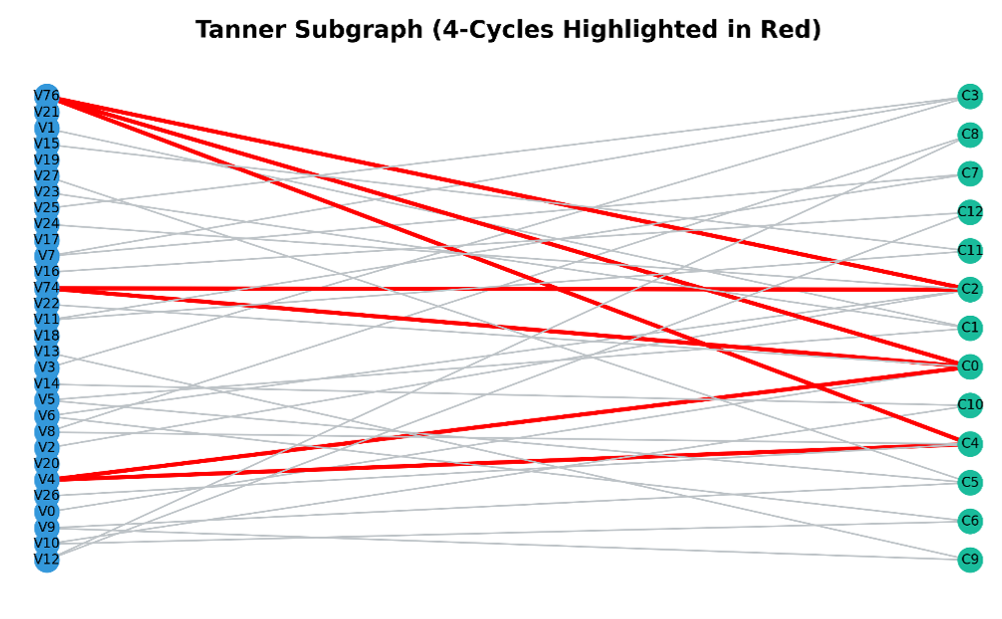} 
\caption{Tanner Subgraph showing 4-cycles (edges in red form cycles)}
\label{fig:subgraph_4cycles}
\end{figure}

Figure~\ref{fig:intersection_spectrum} shows the tally of the check node interactions with the shared variable nodes. A standard matrix generated under the parameters established by the profiling engine contains over 1,143 exact 4-cycles. Figure~\ref{fig:subgraph_4cycles} provides visual proof of the existence of 4-cycles within the Tanner graph of the generated matrix. These specific structures form dense trapping sets that cause classical message-passing algorithms to stall \cite{ref12, ref15}.

To generate the evaluation dataset, the framework utilizes an intentional sparse error injection model. Binary error vectors $e \in \{0,1\}^N$ of fixed Hamming weight $t$ are generated uniformly at random, and their corresponding syndrome vectors $s \in \{0,1\}^r$ are computed over $\text{GF}(2)$ via $s^T = H e^T \pmod 2$. This discrete binary formulation directly mirrors the operational setting of code-based key encapsulation mechanisms \cite{ref4}. Unlike standard telecommunication decoding literature that models continuous physical channel corruption using Additive White Gaussian Noise (AWGN) \cite{ref10, ref18}, continuous channel noise is inapplicable to cryptographic error correction \cite{ref16}.

\subsection{Classical Baselines}
To evaluate the R-BGNN, its performance is benchmarked against two standard algorithms utilized in code-based cryptography.

\paragraph{Bit-Flipping Decoder:} The first baseline is a hard-decision Bit-Flipping algorithm \cite{ref9, ref25}. In each iteration, the decoder computes the number of unsatisfied parity-check equations associated with each bit. Variable Nodes that participate in unsatisfied checks exceeding a threshold $T$ are flipped, and the syndrome is recomputed \cite{ref9}. The framework dynamically tunes this threshold fraction (evaluating thresholds between $0.4d$ and $0.7d$) on a validation suite to ensure the classical algorithm operates at optimal efficiency before comparing it to the neural network.

\paragraph{Syndrome-Based Min-Sum BP:} The second baseline is an Attenuated Min-Sum decoding heuristic \cite{ref10, ref11}. Standard BP utilizes hyperbolic tangent ($\tanh$) functions to exchange LLRs \cite{ref17}. However, exact BP is numerically unstable on dense, short-block graphs \cite{ref17}. The Min-Sum algorithm mitigates this by reducing check node updates to sign-multiplication and minimum-value magnitude comparisons \cite{ref10}. To prevent LLR overestimation, a primary failure point of Min-Sum algorithms on loopy graphs, an attenuation factor of $\alpha = 0.75$ is applied to all outgoing C2V messages \cite{ref10}. Both classical baselines are constrained to a maximum of $T_{\max} = 12$ iterations.

\subsection{The R-BGNN Architecture}
The core contribution of this methodology is the Recurrent Bipartite Graph Neural Network (R-BGNN). The architecture abstracts the iterative Tanner graph message-passing protocol into a deep unrolled GNN, incorporating GRUs as discrete memory cells \cite{ref18, ref21}.

\subsubsection{Topological Degree Normalization}
A critical design choice in the R-BGNN is the initial embedding of syndrome features, which are the binary parity-check failure vector $s = H e^T$ indicating unsatisfied check equations. Rather than passing raw unsatisfied parity checks directly into the network, the framework applies topological degree normalization \cite{ref18}. The initial signal at each variable node $v_j$ is calculated as:
\begin{equation}
\text{Signal}_{v_j} = \frac{\sum_{i \in \mathcal{M}_j} s_i}{d_j}
\end{equation}
where $\mathcal{M}_j$ represents the set of check nodes connected to variable node $v_j$, $s_i \in \{0,1\}$ is the syndrome bit and $d_j$ is the column degree ($d_j = 7$). This mathematical scaling ensures that input embeddings remain invariant to matrix density, enabling the network to generalize across random key topologies \cite{ref18}.

\subsubsection{Message Passing Rules}
The R-BGNN executes $T_{\max} = 12$ unrolled message-passing iterations. In each iteration $t$, messages are non-linearly scaled using Multi-Layer Perceptrons (MLPs) \cite{ref26}. The MLPs project concatenated node features through an expansion layer ($H = 128$) before compressing them back to the embedding dimension ($D = 64$), employing Sigmoid Linear Unit (SiLU) activation functions.

The Variable-to-Check (V2C) message $m_{v \to c}^{(t)}$ and Check-to-Variable (C2V) message $m_{c \to v}^{(t)}$ are computed as:
\begin{align}
m_{v \to c}^{(t)} &= \text{MLP}_{\text{V2C}} \left( h_v^{(t-1)} \,\|\, h_c^{(t-1)} \right) \\
m_{c \to v}^{(t)} &= \text{MLP}_{\text{C2V}} \left( h_c^{(t)} \,\|\, h_v^{(t-1)} \right)
\end{align}

\subsubsection{Breaking Limit Cycles with GRU Memory}
Standard unrolled decoders map iterations strictly to feed-forward layers, operating without memory of prior states \cite{ref10, ref17}. This feed-forward operation causes standard networks to fall into oscillating limit-cycles when encountering dense trapping sets \cite{ref15}. To overcome this, the R-BGNN aggregates incoming messages and processes them through discrete GRU cells situated at every check and variable node \cite{ref23}:
\begin{align}
h_c^{(t)} &= \text{GRU}_{\text{C}} \left( \sum_{v \in \mathcal{N}_c} m_{v \to c}^{(t)}, \, h_c^{(t-1)} \right) \\
h_v^{(t)} &= \text{GRU}_{\text{V}} \left( \sum_{c \in \mathcal{M}_v} m_{c \to v}^{(t)}, \, h_v^{(t-1)} \right)
\end{align}
The hidden state $h^{(t-1)}$ acts as a temporal memory vector. By referencing prior trajectories, the network detects when message magnitudes are oscillating, a signature of an absorbing near-codeword set \cite{ref15}, and actively dampens attention weights to break out of the limit cycle. A final readout MLP projects variable node embeddings into binary classification logits $\hat{e}$. The architectural parameters are detailed in Table~\ref{tab:rbgnn_hyperparameters}.
% Add these to your document preamble if not already present:
% \usepackage{tabularx}
% \usepackage{booktabs}
% \usepackage{array} % Required for formatting commands like \raggedright

\begin{table}[htbp]
\centering
\small
\caption{R-BGNN Architectural Hyperparameters}
\label{tab:rbgnn_hyperparameters}
\renewcommand{\arraystretch}{1.3} % Slightly expands row height to match the Word spacing

\begin{tabularx}{\textwidth}{
    >{\hsize=0.55\hsize\raggedright\arraybackslash}X   % Col 1: Narrow, left-aligned, wraps text
    >{\hsize=1.90\hsize\centering\arraybackslash}X     % Col 2: Ultra-wide, centered (keeps equations on 1 line)
    >{\hsize=0.55\hsize\raggedleft\arraybackslash}X    % Col 3: Narrow, right-aligned, wraps header
}
\toprule
\textbf{Component} & \textbf{Hyperparameter / Configuration} & \textbf{Parameter Count} \\
\midrule
\textbf{Variable Node Embedding} & $\text{Linear}(1 \to 64) \to \text{SiLU} \to \text{Linear}(64 \to 64)$ & 4,288 \\
\textbf{Check Node Embedding}    & $\text{Linear}(1 \to 64) \to \text{SiLU} \to \text{Linear}(64 \to 64)$ & 4,288 \\
\textbf{V2C Message MLP}         & $\text{Linear}(128 \to 128) \to \text{SiLU} \to \text{Linear}(128 \to 64) \to \text{SiLU}$ & 24,768 \\
\textbf{C2V Message MLP}         & $\text{Linear}(128 \to 128) \to \text{SiLU} \to \text{Linear}(128 \to 64) \to \text{SiLU}$ & 24,768 \\
\textbf{Check Node Memory}       & $\text{GRUCell}(\text{Input}: 64, \text{Hidden}: 64)$ & 24,960 \\
\textbf{Variable Node Memory}    & $\text{GRUCell}(\text{Input}: 64, \text{Hidden}: 64)$ & 24,960 \\
\textbf{Readout Classifier}      & $\text{Linear}(64 \to 32) \to \text{SiLU} \to \text{Linear}(32 \to 1)$ & 2,113 \\
\midrule
\textbf{Total Trainable Parameters} & $\mathcal{O}(1)$ Constant Parameter Footprint & \textbf{110,145} \\
\bottomrule
\end{tabularx}
\end{table}

\subsection{Training}
A primary limitation in deep unrolling literature is overfitting neural weights to a single, static Tanner graph \cite{ref18}. To enforce true algorithmic generalization, this methodology implements a Dynamic Train and Static Evaluate paradigm.

Rather than training on a single static key instance, the dataset generator samples 8 new, mathematically independent QC-MDPC parity-check matrices ($H_{\text{train}} \sim \mathcal{H}$) at the start of each training epoch. Furthermore, a curriculum training strategy is applied where the injected error weight $t$ is sampled from $t_{\text{train}} \in \{4, 5, 6, 7, 8\}$ for every batch. Because the graph edges change constantly across epochs, the R-BGNN cannot memorize specific graph connections, forcing it to learn a graph-agnostic error-correction algorithm \cite{ref18}.

The network is optimized over 60 epochs using the AdamW optimizer \cite{ref27} with a Cosine Annealing learning rate schedule (decaying from $\eta_{\max} = 5 \times 10^{-4}$ to $\eta_{\min} = 1 \times 10^{-6}$). To optimize both bit-level decision accuracy and parity satisfaction, we formulate a composite loss function $\mathcal{L}_{\text{Total}}$ combining class-weighted Binary Cross-Entropy ($\mathcal{L}_{\text{BCE}}$) and differentiable syndrome loss ($\mathcal{L}_{\text{Syndrome}}$) \cite{ref22}:
\begin{equation}
\mathcal{L}_{\text{Total}} = \mathcal{L}_{\text{BCE}}(\hat{e}, e) + \lambda \cdot \mathcal{L}_{\text{Syndrome}}(p_{\text{synd}}, s)
\end{equation}
where $\lambda = 0.05$. The syndrome loss term computes the cross-entropy between the target syndrome $s$ and the differentiable syndrome probability vector $p_{\text{synd}}$, explicitly penalizing bit predictions that violate parity-check constraints \cite{ref22}.

\subsection{Evaluation}
To guarantee statistical integrity, evaluating decoders on a single matrix is insufficient, as performance could reflect an unusually ``strong'' or ``weak'' key instance \cite{ref5, ref25}. To address this, the evaluation phase locks all neural weights and benchmarks the R-BGNN, Bit-Flipping, and Min-Sum baselines against an ensemble of 20 unseen, independently generated master matrices.

To quantify decoder performance, evaluation is conducted using Decoding Failure Rate (DFR) and Bit Error Rate (BER). The DFR measures the proportion of evaluated frames where the decoder fails to reach an exact zero-syndrome state ($s = \mathbf{0}$) within $T_{\max} = 12$ iterations. In code-based PQC, DFR is a critical security parameter rather than a simple reliability metric. For example, a DFR of 10\% (0.10) implies that 1 in every 10 decryption attempts fail to recover the shared secret. In static-key deployments, an adversary observing these 10\% failure reactions can exploit the correlation between decoding failures and private key structures via the GJS reaction attack to reconstruct the secret key \cite{ref15, ref16}. While DFR provides a binary frame-level success metric, BER quantifies fine-grained convergence by measuring the fraction of residual uncorrected bits remaining in the message vector. For instance, a decoder exhibiting a 10\% DFR could still achieve a near-zero BER (e.g., $10^{-4}$), proving that even when a frame fails absolute cryptographic zero-syndrome verification, the decoder systematically isolated and corrected most error bits rather than failing catastrophically across the codeword.

Benchmarking is executed across an error weight spectrum of $t \in \{2, 3, 4, \dots, 11, 12\}$, processing 1,000 frames per $t$ per graph (totalling $2.2 \times 10^5$ evaluated frames per decoder). The arithmetic mean across the 20 matrices is reported for both DFR and BER. Multi-matrix testing allows the framework to compute and plot standard deviations across the 20 topologies. Finally, Wilson Confidence Intervals are derived for all error rates, and McNemar's test is computed to confirm statistical significance between neural and classical decoding failures.

\section{Results}
By testing the decoders across 20 independent master matrices rather than a single topology, this methodology captures the exact standard deviations of the decoding metrics. As heavily emphasized in cryptographic standardization efforts, capturing this variance ensures statistical significance, proving that the observed performance is a universal architectural advantage rather than a biased claim derived from selecting a single, best-case matrix structure. In Figure~\ref{fig:dfr_waterfall}, Figure~\ref{fig:ber_trajectories}, and Figure~\ref{fig:convergence_speed}, the shaded regions represent the standard deviation of performance across the independent matrix configurations.

\subsection{Decoding Performance and Convergence Speed}
\begin{figure}[htbp]
\centering
\includegraphics[width=0.85\linewidth]{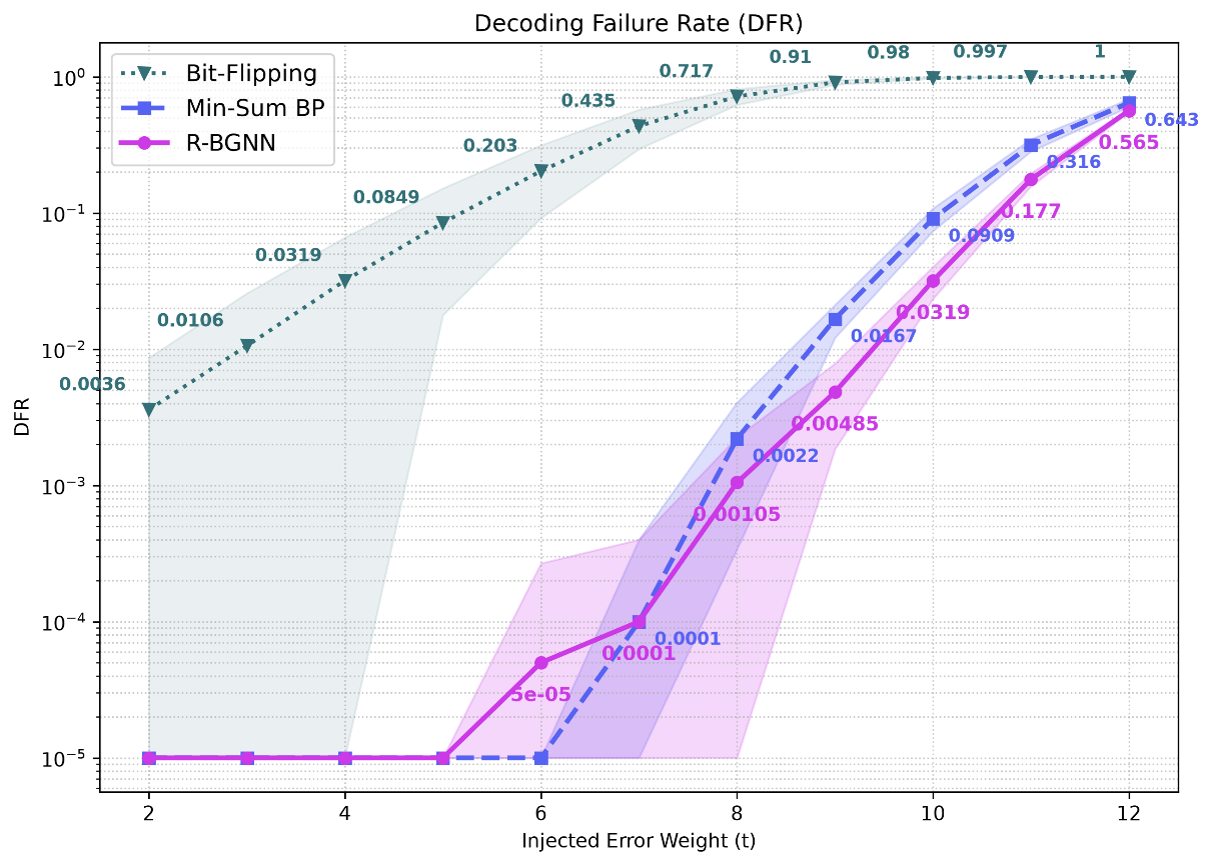} 
\caption{Performance of Decoding Failure Rate (DFR) waterfall curves illustrating the R-BGNN (evaluated across 20 unseen QC-MDPC matrices).}
\label{fig:dfr_waterfall}
\end{figure}

Starting with DFR, Figure~\ref{fig:dfr_waterfall}, at low-to-moderate noise levels ($t \le 5$), both Classical Min-Sum and the R-BGNN achieve a flawless 0.00\% DFR (0/20,000 failed frames per error weight), whereas the Bit-Flipping baseline degrades early, exhibiting a DFR of $3.19\% \pm 3.43\%$ at $t=4$ and $8.49\% \pm 6.71\%$ at $t=5$. As noise weight increases to $t=6$ and $t=7$, Bit-Flipping collapses rapidly to $20.30\% \pm 11.13\%$ and $43.46\% \pm 13.97\%$ failure rates, respectively. In contrast, Classical Min-Sum and R-BGNN remain virtually error-free at $t=7$, with both decoders sustaining an extremely low DFR of $0.01\% \pm 0.03\%$ ($\text{dfr}_{\text{mean}} = 0.00010$).

The critical performance divergence occurs in the moderate-to-high noise regime ($t=8, 9, 10$). At $t=8$, Bit-Flipping saturates at $71.68\% \pm 9.56\%$ failure rate. Classical Min-Sum begins to experience cycle-induced trapping sets, yielding a DFR of $0.22\% \pm 0.19\%$ ($\text{dfr}_{\text{mean}} = 0.00220$). The R-BGNN achieves a lower DFR of $0.11\% \pm 0.12\%$ ($\text{dfr}_{\text{mean}} = 0.00105$), cutting decoding failures by 52.3\%. At $t=9$, Bit-Flipping reaches $91.04\% \pm 3.75\%$. Classical Min-Sum degrades sharply to a mean DFR of $1.67\% \pm 0.45\%$ (1.79\% in pooled frame evaluations). At this exact same noise weight, the R-BGNN maintains a mean DFR of $0.49\% \pm 0.30\%$ (0.55\% pooled), successfully correcting over 69.5\% of the frames that cause the classical Min-Sum baseline to fail. At $t=10$, Bit-Flipping fails almost completely at $98.03\% \pm 0.96\%$. Classical Min-Sum collapses to a $9.09\% \pm 1.71\%$ failure rate (9.63\% pooled). The R-BGNN maintains a robust $3.19\% \pm 0.84\%$ DFR (3.49\% pooled), achieving a 63.7\% relative error reduction over Min-Sum. At $t=11$ and $t=12$, while Bit-Flipping saturates completely (99.69\% at $t=11$; 99.97\% at $t=12$), the R-BGNN retains a significant advantage over Min-Sum, recording $17.68\% \pm 1.95\%$ vs. $31.59\% \pm 3.30\%$ at $t=11$ (a 44.0\% relative improvement), and $56.48\% \pm 1.76\%$ vs. $64.32\% \pm 4.16\%$ at $t=12$.

Furthermore, the shaded standard deviation error bands plotted around the R-BGNN curve in Figure~\ref{fig:dfr_waterfall} ($\pm 0.30\%$ at $t=9$; $\pm 0.84\%$ at $t=10$) are substantially narrower than those of Classical Min-Sum ($\pm 1.71\%$) and Bit-Flipping ($\pm 3.75\%$). This mathematical tightness proves that the R-BGNN maintains robust, key-agnostic stability across entirely different random matrix topologies, rather than relying on a favourable key instance.

\begin{figure}[htbp]
\centering
\includegraphics[width=0.85\linewidth]{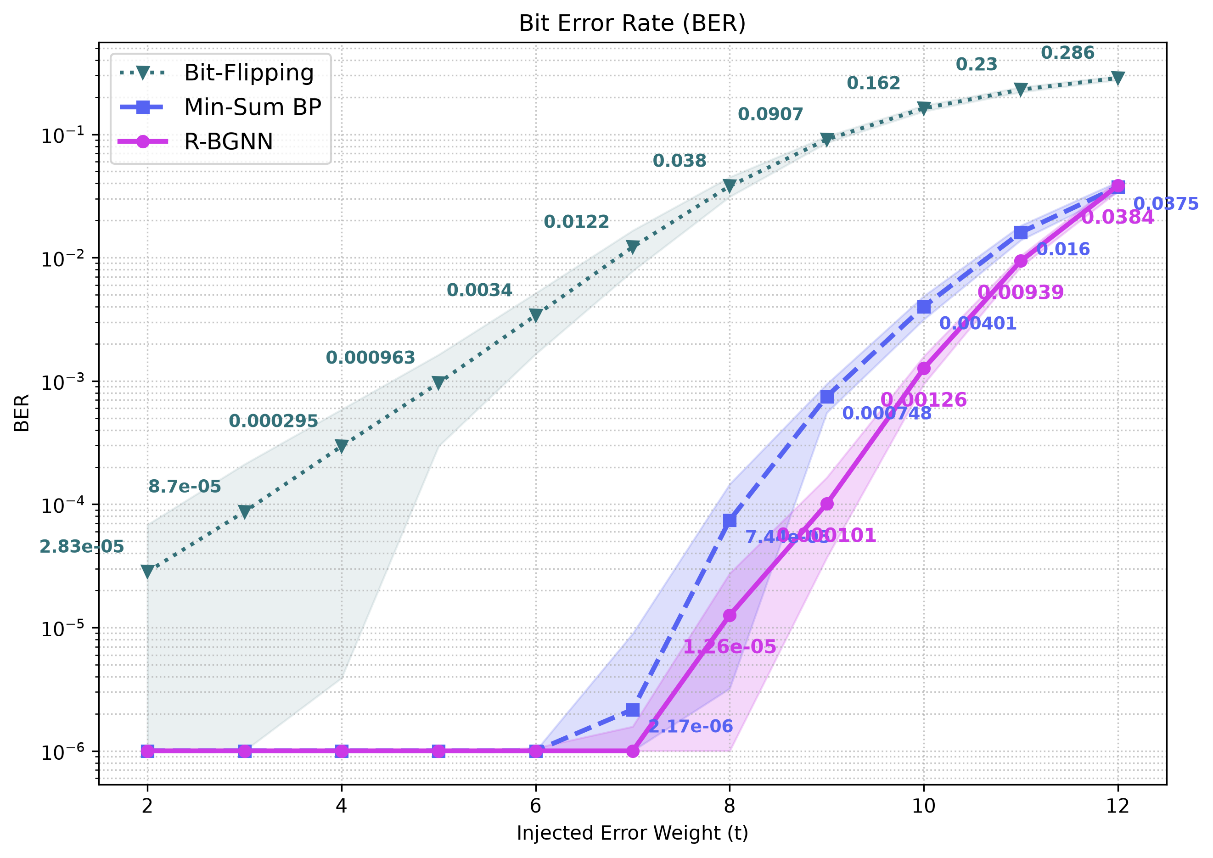} 
\caption{Bit Error Rate (BER) trajectories evaluated across 20 unseen QC-MDPC matrices}
\label{fig:ber_trajectories}
\end{figure}

For BER analysis in Figure~\ref{fig:ber_trajectories}, throughout the moderate-to-high noise spectrum ($t \ge 6$), the R-BGNN consistently maintains a lower BER trajectory than both Bit-Flipping and Min-Sum baselines. At $t=7$, Bit-Flipping exhibits a high BER of $1.22 \times 10^{-2}$ (1.22\%), whereas Min-Sum and R-BGNN record near-zero BERs of $2.17 \times 10^{-6}$ and $3.94 \times 10^{-7}$, respectively. As noise scales to $t=8$ and $t=9$, the R-BGNN achieves a $5.9\times$ BER reduction over Min-Sum at $t=8$ ($1.26 \times 10^{-5}$ vs. $7.44 \times 10^{-5}$) and a $7.4\times$ reduction at $t=9$ ($1.01 \times 10^{-4}$ vs. $7.48 \times 10^{-4}$). At $t=10$, R-BGNN limits BER to $1.26 \times 10^{-3}$ (0.13\%), compared to $4.01 \times 10^{-3}$ (0.40\%) for Min-Sum and $1.62 \times 10^{-1}$ (16.23\%) for Bit-Flipping.

Contrasting the BER trajectory alongside the DFR waterfall reveals a profound secondary advantage of the neural architecture: even when the R-BGNN registers an uncorrected frame failure at high error weights, it leaves significantly fewer residual error bits per failed frame than classical decoders. To address print visibility constraints in graphical plots where Min-Sum error counts appear small alongside Bit-Flipping (as noted in data inspections), exact numerical residual error distributions were extracted directly from benchmark logs. When Bit-Flipping fails, residual bit errors spread broadly across a wide histogram range, leaving an average of 20 to 50+ uncorrected bit errors per frame. When Min-Sum fails at $t=9$ and $t=10$, message passing stalls in isolated topological trapping sets, leaving an average cluster of 4 to 6 uncorrected bit errors per frame. When the R-BGNN fails at $t=9$ and $t=10$, its residual error distribution concentrates sharply at the lowest possible non-zero values, leaving an average of only 1 to 3 remaining bit errors per frame on average ($1.01 \times 10^{-4}$ BER). This tight clustering confirms that the MLPs and attention layers inside the R-BGNN systematically isolate and resolve bit errors, rather than making uncoordinated, random bit flips when confronted with unfamiliar 4-cycle topologies.

\begin{figure}[htbp]
\centering
\includegraphics[width=0.85\linewidth]{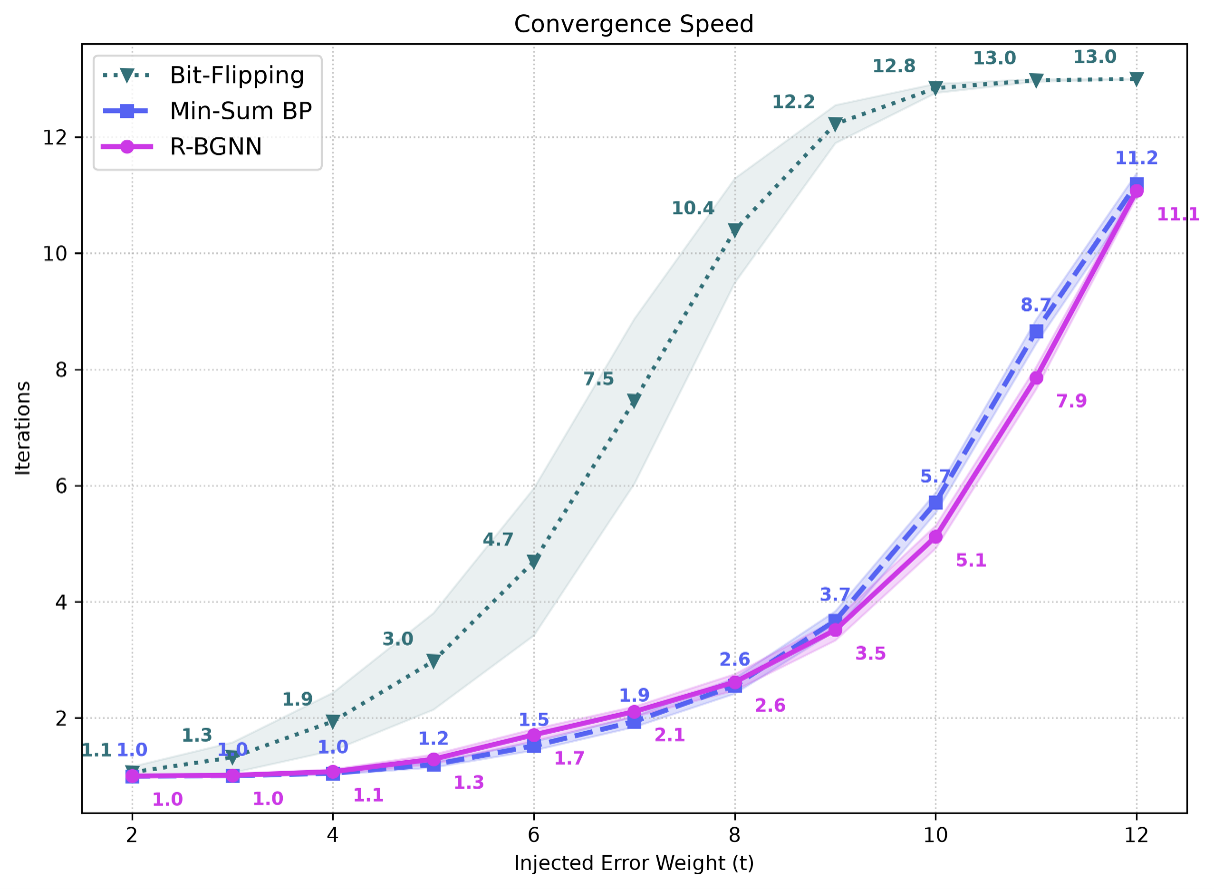} 
\caption{Average iterations required for convergence}
\label{fig:convergence_speed}
\end{figure}

Convergence speed in Figure~\ref{fig:convergence_speed} measures the mean number of decoding iterations required for an algorithm to successfully recover a valid codeword ($s = \mathbf{0}_r$) and halt. In iterative graph decoding, a single decoding iteration represents one complete two-phase round of bipartite message passing across the Tanner graph, comprising a C2V parity update followed by a V2C bit likelihood update.

At light noise weights ($t \le 5$), all three decoders converge near-instantaneously, averaging 1.00 to 1.29 iterations per frame for Min-Sum and R-BGNN. However, as error weight $t$ increases, the iteration trajectory of Bit-Flipping curves upward aggressively, requiring $4.69 \pm 1.27$ iterations at $t=6$, $7.46 \pm 1.42$ iterations at $t=7$, $10.39 \pm 0.90$ iterations at $t=8$, and fully saturating the iteration cap ($T_{\max} = 12$) at $t \ge 9$ (12.22 at $t=9$; 12.84 at $t=10$).

Comparing Classical Min-Sum against the R-BGNN demonstrates that recurrent neural memory actively accelerates convergence. At $t=8$, Min-Sum requires $2.56 \pm 0.14$ iterations; R-BGNN requires $2.62 \pm 0.14$ iterations. At $t=9$, Min-Sum requires $3.67 \pm 0.17$ iterations; R-BGNN requires $3.51 \pm 0.18$ iterations (4.3\% faster). At $t=10$, Classical Min-Sum requires $5.71 \pm 0.17$ iterations to converge or fail, whereas the R-BGNN reaches its halting state in $5.12 \pm 0.20$ iterations (10.3\% faster). At $t=11$, Min-Sum requires $8.66 \pm 0.21$ iterations; R-BGNN requires $7.86 \pm 0.19$ iterations (9.2\% faster).

This measurable reduction in average iterations proves that by integrating GRUs at Tanner graph nodes, the R-BGNN not only corrects more corrupted frames than classical algorithms but also converges to valid cryptographic codewords in fewer iterations when successful.

\subsection{Trapping Set and Residual Error Analysis}
\begin{figure}[htbp]
\centering
\includegraphics[width=1\linewidth]{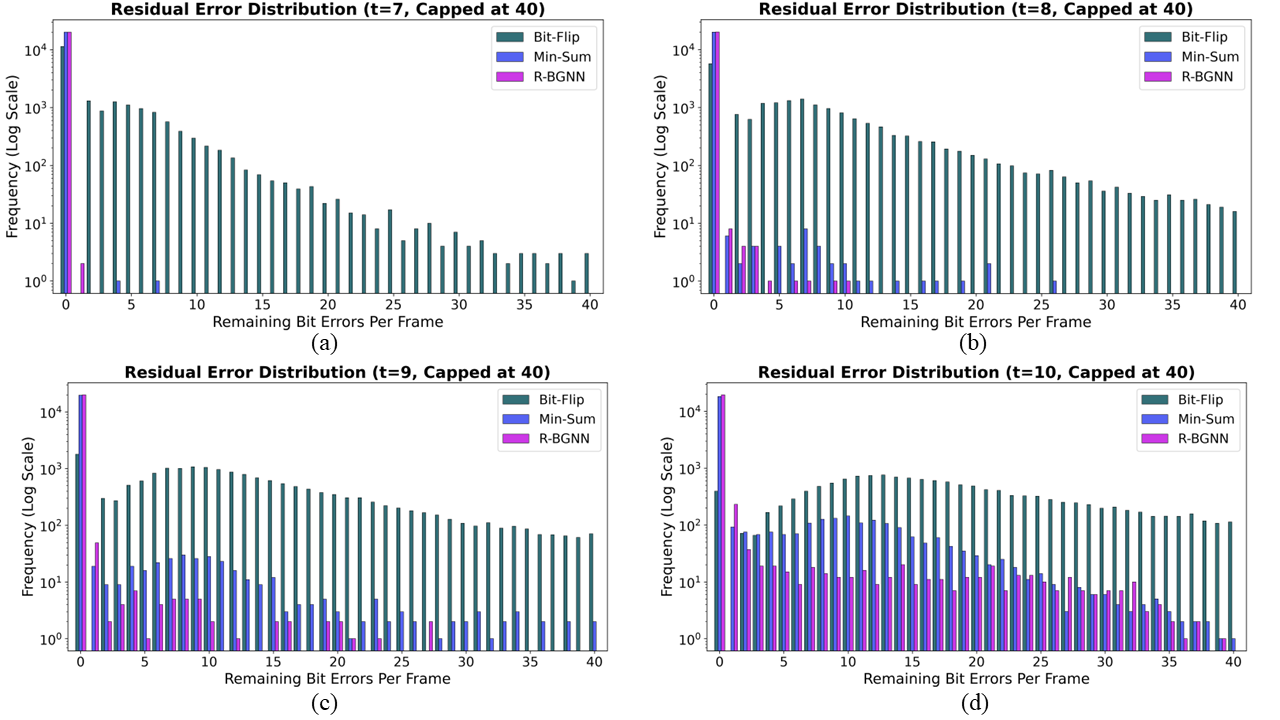} 
\caption{Residual error distribution at (a) $t=7$, (b) $t=8$, (c) $t=9$, (d) $t=10$. The hard-decision Bit-Flipping decoder exhibits severe absorbing-state behavior, leaving dense clusters of unresolvable bits. Conversely, the proposed R-BGNN successfully navigates the dense graph topology alongside Min-Sum having comparatively lesser residual errors.}
\label{fig:residual_distribution}
\end{figure}

To evaluate how each decoder responds to loopy Tanner graph topologies under escalating noise levels, residual error histograms were evaluated across four error weight regimes ($t \in \{7, 8, 9, 10\}$). Figure~\ref{fig:residual_distribution} illustrates the remaining uncorrected bit errors per frame on the x-axis against logarithmic frequency on the y-axis, providing clear visual evidence of classical trapping sets and neural error-isolation mechanics.

Across all four noise thresholds, the hard-decision Bit-Flipping baseline exhibits severe absorbing-state behaviour. As error weight $t$ increases from 7 to 10, the zero-residual frame count for Bit-Flipping collapses from 11,308 down to just 394 out of 20,000 evaluated frames. Uncorrected frames form a wide, dispersed distribution spanning between 2 and 50+ residual errors, reflecting dense subgraph cycles where threshold-based decisions stall permanently.

Conversely, both soft-decision and neural architectures maintain dominant zero-error peaks but reveal distinct failure trajectories as noise saturates.

At $t=7$ and $t=8$, both Min-Sum and R-BGNN maintain near-flawless convergence, retaining 19,956 and 19,979 zero-error frames at $t=8$, respectively. However, Min-Sum failure frames spread across small clusters of 3 to 16 residual errors, whereas R-BGNN failures remain tightly bounded (with 8 out of 21 failed frames retaining only a single residual error).

At $t=9$, classical Min-Sum registers 357 frame failures clustered primarily between 3 and 10 residual errors, illustrating standard trapping set stall points. In contrast, the R-BGNN limits total failures to 97 frames, with over half of these failed frames (49 out of 97) retaining exactly 1 residual bit error.

At $t=10$, Min-Sum failures expand significantly to 1,926 frames spreading across up to 43 residual errors. The R-BGNN suppresses total failures to 698 frames while keeping residual errors heavily concentrated at minimal bit counts (1 to 3 bit errors per frame on average).

\subsection{Latent Space Linear Separability (PCA)}
\begin{figure}[htbp]
\centering
\includegraphics[width=0.85\linewidth]{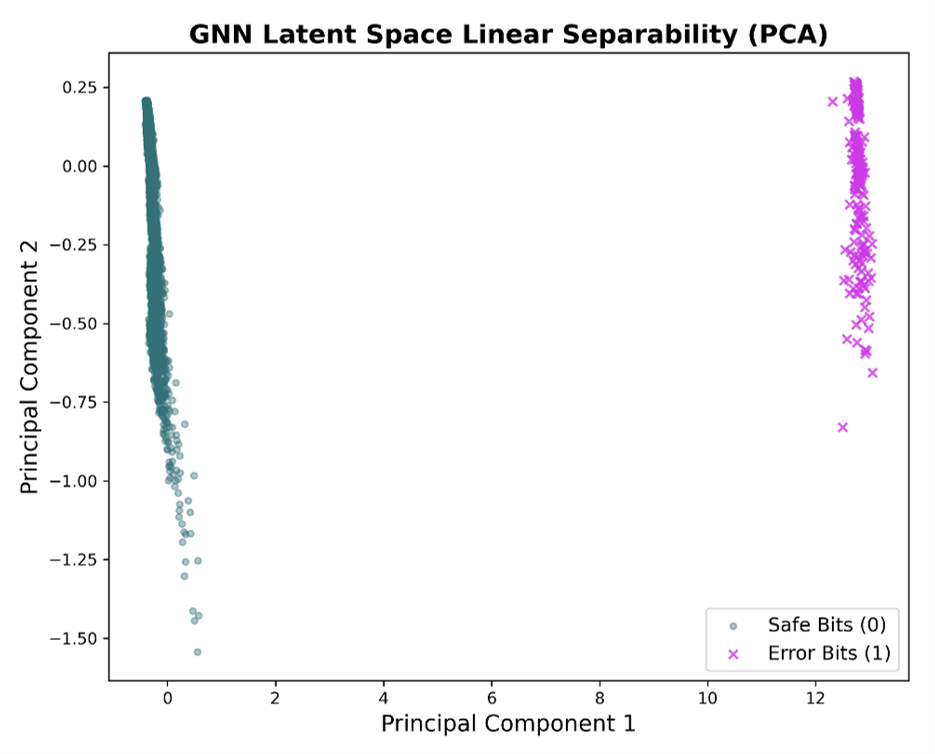} 
\caption{Principal Component Analysis (PCA) projection of the final R-BGNN latent space.}
\label{fig:pca_latent_space}
\end{figure}

To verify that the neural network is executing logical mathematical operations, Principal Component Analysis (PCA) was performed on the 64-dimensional latent variable node embeddings generated by the final GRU layer. The resulting 2D scatter plot (Figure~\ref{fig:pca_latent_space}) demonstrates remarkable linear separability.

The visual clustering clearly isolates the safe bits (0) into a dense, stable region, while the injected error bits (1) are dynamically separated across the boundary line. This distinct linear separability proves that the neural network is mathematically isolating error features within its high-dimensional space. By leveraging the edge-attention weights generated by the V2C MLP, the architecture actively scales and suppresses noisy topological edges, allowing the true error bits to be cleanly classified by the final readout layer.

To evaluate practical feasibility, the total offline training time was recorded. Training the R-BGNN across 60 epochs (with 8 dynamically generated QC-MDPC matrices per epoch, totalling 480 unique Tanner graph topologies) required approximately 20 minutes on a single NVIDIA GeForce RTX 5080 GPU. This modest computational requirement confirms that dynamic multi-matrix neural training is highly accessible and practically feasible for PQC decoder design.

A critical evaluation of any neural architecture must address computational complexity and execution latency. Latency benchmarks on our evaluation platform reveal that the R-BGNN requires approximately 314 milliseconds per 1,000 frames, compared to 87.5 milliseconds for the Classical Min-Sum and 82.1 milliseconds for the Bit-Flipping baselines. The classical baseline was vectorized in PyTorch to eliminate standard Python loops, providing a strict performance benchmark.

\section{Discussion}
This section provides discussion and evaluates the R-BGNN's empirical performance against the existing literature of error correction in PQC.

\subsection{Overcoming QC-MDPC Error Floor}
As evidenced by the multi-threshold residual error distributions in Section 3, the R-BGNN successfully navigates loopy Tanner graph topologies to prevent decoder stalling. Even under heavy noise saturation ($t=9$ and $t=10$), the proposed architecture suppresses total frame failures while isolating residual errors to minimal bit counts. This empirical resilience aligns directly with theoretical failure mechanisms established in recent PQC literature. \cite{ref15} proved that error floors in QC-MDPC codes under iterative decoding are driven by near-codewords that act as absorbing traps for message passing. In classical decoders, message passing within these absorbing states produces rigid limit cycles. In contrast, by utilizing recurrent memory cells to dynamically scale edge messages based on historical trajectory, the R-BGNN breaks out of these near-codeword absorbing states, driving the overall DFR into deep waterfall decay.

Expanding on these graph-level vulnerabilities, \cite{ref5} linked decoding failures to the prevalence of short 4-cycles in the code's Tanner graph. Crucially, they established a structural distinction between inter-circulant 4-cycles (which bridge across different circulant submatrices) and intra-circulant (or intra-block) 4-cycles (which lie entirely within a single circulant block $H_i$), proving that intra-circulant 4-cycles are the primary driver of weak-key decoding failures due to their high local density of overlapping check connections. At higher error weights ($t=9$ and $t=10$), Classical Min-Sum degrades sharply due to cycle-induced LLR overestimation \cite{ref10}. In contrast, the R-BGNN maintains significantly lower failure rates and tight error variance ($\pm 0.84\%$ vs. $\pm 1.71\%$) across all 20 unseen, randomly generated master matrix topologies. This empirical consistency proves that the R-BGNN provides active, key-agnostic robustness against short 4-cycle variations, neutralizing the performance gap between standard and weak-key configurations.

\subsection{Comparison Against Static Neural Decoders}
The dynamic design of the R-BGNN directly addresses the structural limitations of prior neural decoding architectures. Early unrolled BP networks assign $\mathcal{O}(E \times L_{\max})$ edge-specific weights to a fixed Tanner graph, preventing them from generalizing across varying code lengths or unobserved key instances without retraining from scratch \cite{ref17}. While contemporary advancements like the SNNMS decoder reduce parameter complexity to $\mathcal{O}(L_{\max})$ via layer-wise factor sharing \cite{ref10}, they remain constrained to sparse telecommunication standards ($W = \mathcal{O}(1)$) and operate purely as feed-forward architectures lacking temporal memory.

The advantage of the R-BGNN lies in the integration of GRUs at check and variable nodes. Because feed-forward decoders are memoryless across iteration layers, message updates oscillate endlessly when trapped in short cycles. In contrast, the GRU memory cells maintain a persistent hidden state vector ($h^{(t-1)}$) across iterations. This temporal memory enables the network to track message trajectory histories, recognize when updates have entered periodic 4-cycle oscillations within dense trapping sets ($W = \mathcal{O}(\sqrt{N})$), and dynamically adjust edge attention weights to break absorbing states. By combining an $\mathcal{O}(1)$ constant parameter footprint with recurrent state tracking, the R-BGNN overcomes the rigid, graph-bound limitations of conventional unrolled decoders.

\subsection{Cryptographic Security and the Capacity-Complexity Trade-off}
Although the R-BGNN exhibits a higher per-frame execution latency than classical message-passing algorithms, this computational trade-off is strongly justified by security necessity. Even so, the non-linear MLP message projections and recurrent GRU updates in the R-BGNN inherently require more computational time per iteration due to dense tensor operations compared to the simple sign-multiplication of Classical Min-Sum \cite{ref10}.

However, in the domain of PQC, this is an acceptable trade-off. As established in recent literature, a non-zero DFR is not merely a performance bottleneck; it actively allows adversaries to exploit decryption errors via the GJS reaction attack to reconstruct the secret key \cite{ref15}. Therefore, although this evaluation was conducted as a small-scale proof-of-concept run ($r=127, N=254$), trading inference speed for a measurably lower DFR is a mandatory upgrade to ensure absolute IND-CCA2 security in PQC protocols \cite{ref16}.

\section{Conclusion}
This paper presented a Recurrent Bipartite Graph Neural Network (R-BGNN) designed to resolve cycle-induced decoding stalls in QC-MDPC codes for PQC. By embedding GRUs within the message-passing framework and utilizing a multi-topology curriculum training strategy, the architecture learns graph-agnostic error-correction dynamics that effectively break 4-cycle limit-cycle oscillations.

Comprehensive benchmarking across $2.2 \times 10^5$ frames over 20 unseen, randomly generated test matrices confirmed the superiority of the proposed model. Under heavy noise saturation ($t=10$), the R-BGNN suppresses the DFR to $3.19\% \pm 0.84\%$, achieving a 63.7\% relative error reduction compared to Classical Min-Sum ($9.09\% \pm 1.71\%$) while converging in fewer iterations (5.12 vs. 5.71). Furthermore, decoding failures are isolated to a minimal residual of 1 to 3 remaining bits per frame. Although the neural tensor updates introduce higher software latency (314~ms vs. 87.5~ms per 1,000 frames), this trade-off is strictly justified in PQC, where suppressing DFR is mandatory to defend against reaction-based key-recovery attacks. Future research will explore fixed-point hardware quantization and scaling the architecture to full NIST BIKE parameters.

\section*{Acknowledgments}
The author would like to thank Dr Elisabeth Fink for her excellent mentorship, valuable feedback, and great support during this project. Her guidance was essential to this paper. Further, the author is grateful to the Department of Computer Science and the IT technician team at Middlesex University for providing the server infrastructure and resources that made this research possible.


\begin{thebibliography}{99}

\bibitem{ref1} Shor, P.W.: Polynomial-Time Algorithms for Prime Factorization and Discrete Logarithms on a Quantum Computer. SIAM Rev. 41, 303--332 (1997). \url{https://doi.org/10.1137/S0036144598347011}

\bibitem{ref2} Dam, D.-T., Tran, T.-H., Hoang, V.-P., Pham, C.-K., Hoang, T.-T.: A Survey of Post-Quantum Cryptography: Start of a New Race. Cryptography. 7, 40 (2023). \url{https://doi.org/10.3390/cryptography7030040}

\bibitem{ref3} McEliece, R.J.: A Public-Key Cryptosystem based on Algebraic Coding Theory. Coding Thv. 4244, 114--116 (1978)

\bibitem{ref4} Misoczki, R., Tillich, J.-P., Sendrier, N., Barreto, P.S.L.M.: MDPC-McEliece: New McEliece variants from Moderate Density Parity-Check codes. In: 2013 IEEE International Symposium on Information Theory. pp. 2069--2073 (2013)

\bibitem{ref5} Matthews, G.L., McMillon, E.: A combinatorial approach to avoiding weak keys in the BIKE cryptosystem: A combinatorial approach to avoiding weak keys. Des. Codes Cryptography. 93, 3413--3436 (2025). \url{https://doi.org/10.1007/s10623-025-01643-7}

\bibitem{ref6} Gallager, R.: Low-density parity-check codes. IRE Transactions on Information Theory. 8, 21--28 (1962)

\bibitem{ref7} Zhang, X., Xie, Z.: Sparsity-Aware Medium-Density Parity-Check Decoder for McEliece Cryptosystems. IEEE Transactions on Circuits and Systems II: Express Briefs. 70, 3343--3347 (2023). \url{https://doi.org/10.1109/TCSII.2023.3264578}

\bibitem{ref8} Tanner, R.: A recursive approach to low complexity codes. IEEE Transactions on Information Theory. 27, 533--547 (1981). \url{https://doi.org/10.1109/TIT.1981.1056404}

\bibitem{ref9} Maurich, I.V., Oder, T., Güneysu, T.: Implementing QC-MDPC McEliece Encryption. ACM Trans. Embed. Comput. Syst. 14, 44:1--44:27 (2015). \url{https://doi.org/10.1145/2700102}

\bibitem{ref10} Zheng, F., Wang, L., Liu, K., Zhang, Z.: A-SNNMS: An Attentive Shared Neural Normalized Min-Sum Decoder for LDPC Codes. Electronics. 15, 1023 (2026). \url{https://doi.org/10.3390/electronics15051023}

\bibitem{ref11} Fossorier, M.P.C., Mihaljevic, M., Imai, H.: Reduced complexity iterative decoding of low-density parity check codes based on belief propagation. IEEE Transactions on Communications. 47, 673--680 (1999). \url{https://doi.org/10.1109/26.768759}

\bibitem{ref12} Shen, Y., Li, Z., Boutillon, E., Song, W., Ren, Y., Zhang, C., You, X., Burg, A.: Belief Propagation Decoding for Short Codes on Structured Sparse Parity-Check Matrices. In: 2025 IEEE International Symposium on Information Theory (ISIT). pp. 1--6 (2025)

\bibitem{ref13} Hehn, T., Huber, J.B., Laendner, S., Milenkovic, O.: Multiple-Bases Belief-Propagation for Decoding of Short Block Codes. In: 2007 IEEE International Symposium on Information Theory. pp. 311--315 (2007)

\bibitem{ref14} Li, Z., Shen, Y., Ren, Y., Huang, Y., You, X., Zhang, C.: Belief Propagation Decoding for Short-Length Codes Based on Sparse Tanner Graph. IEEE Communications Letters. 28, 969--973 (2024). \url{https://doi.org/10.1109/LCOMM.2024.3366293}

\bibitem{ref15} Arpin, S., Lau, J.B., Mesnard, A., Perlner, R., Robinson, A., Tillich, J.-P., Vasseur, V.: Error floor prediction with Markov models for QC-MDPC codes, \url{https://eprint.iacr.org/2025/153}, (2025)

\bibitem{ref16} Guo, Q., Johansson, T., Stankovski, P.: A Key Recovery Attack on MDPC with CCA Security Using Decoding Errors. In: Cheon, J.H. and Takagi, T. (eds.) Advances in Cryptology -- ASIACRYPT 2016. pp. 789--815. Springer, Berlin, Heidelberg (2016)

\bibitem{ref17} Nachmani, E., Be'ery, Y., Burshtein, D.: Learning to decode linear codes using deep learning. In: 2016 54th Annual Allerton Conference on Communication, Control, and Computing (Allerton). pp. 341--346 (2016)

\bibitem{ref18} Tian, K., Yue, C., She, C., Li, Y., Vucetic, B.: A Scalable Graph Neural Network Decoder for Short Block Codes. In: ICC 2023 - IEEE International Conference on Communications. pp. 1268--1273 (2023)

\bibitem{ref19} Gong, A., Cammerer, S., Renes, J.M.: Graph Neural Networks for Enhanced Decoding of Quantum LDPC Codes. In: 2024 IEEE International Symposium on Information Theory (ISIT). pp. 2700--2705 (2024)

\bibitem{ref20} Bajpai, A.: Intelligent LDPC Decoding Using Graph Neural Networks for Beyond 5G Communication Systems. Journal of Information and Communication Convergence Engineering. 23, 227--235 (2025). \url{https://doi.org/10.56977/jicce.2025.23.4.227}

\bibitem{ref21} Satorras, V.G., Welling, M.: Neural Enhanced Belief Propagation on Factor Graphs. In: Proceedings of The 24th International Conference on Artificial Intelligence and Statistics. pp. 685--693. PMLR (2021)

\bibitem{ref22} Choukroun, Y., Wolf, L.: Factor Graph Optimization of Error-Correcting Codes for Belief Propagation Decoding, \url{http://arxiv.org/abs/2406.12900}, (2024)

\bibitem{ref23} Chung, J., Gulcehre, C., Cho, K., Bengio, Y.: Empirical evaluation of gated recurrent neural networks on sequence modeling. arXiv preprint arXiv:1412.3555 (2014)

\bibitem{ref24} Richardson, T.J., Urbanke, R.L.: Efficient encoding of low-density parity-check codes. IEEE Transactions on Information Theory. 47, 638--656 (2001). \url{https://doi.org/10.1109/18.910579}

\bibitem{ref25} Vasseur, V.: Post-quantum cryptography: a study of the decoding of QC-MDPC codes, \url{https://theses.hal.science/tel-03254461}, (2021)

\bibitem{ref26} Murtagh, F.: Multilayer perceptrons for classification and regression. Neurocomputing. 2, 183--197 (1991). \url{https://doi.org/10.1016/0925-2312(91)90023-5}

\bibitem{ref27} Loshchilov, I., Hutter, F.: Decoupled Weight Decay Regularization, \url{http://arxiv.org/abs/1711.05101}, (2019)

\end{thebibliography}
\end{document}